\documentclass[preprint,amsmath,amssymb,aps]{revtex4-1}
\usepackage{graphicx}
\usepackage{dcolumn}
\usepackage{bm}
\usepackage{textcomp}
\usepackage{float}
\usepackage[colorlinks]{hyperref}
\usepackage{epstopdf}

\begin{document}

\title{Terahertz detection using mechanical resonators based on 2D materials}

\author{Juha Hassel$^2$, Mika Oksanen$^1$, Teemu Elo$^1$, Heikki Sepp\"a$^2 $ and Pertti J. Hakonen$^1$}

\affiliation{$^1$Low Temperature Laboratory, Department of Applied Physics, Aalto University School of Science, FI-00076, Finland
}
\affiliation{$^2$VTT Technical Research Centre of Finland, PO BOX 1000, FI-02044 VTT, Finland}

\date{\today}

\begin{abstract}
We have investigated a THz detection scheme based on mixing of electrical signals in a voltage-dependent capacitance made out of suspended graphene. We have analyzed both coherent and incoherent detection regimes and compared their performance with the state of the art. Using a high-amplitude local oscillator, we anticipate potential for quantum limited detection in the coherent mode. The sensitivity stems from the extraordinary mechanical and electrical properties of atomically thin graphene or graphene-related 2D materials.
\end{abstract}

\pacs{PACS numbers: }

\maketitle


\section{Introduction}

Interest in THz detection and imaging technologies is traditionally motivated by astronomy and more recently also by a growing demand for new solutions for enhancing public security.
Passive THz imaging using cryogenic sensor arrays has been successful in fulfilling this demand. At present, incoherent detectors based on transition edge sensors and coherent detectors typically based on SIS mixers  have applications in astronomical imaging \cite{Holland2013,Baryshev2015}. Security applications using a few different approaches have been demonstrated as well  \cite{Grossman2010,Heinz2011,Luukanen2012}.  New solutions with enhanced sensitivity, increased operating temperature, or an increased level of integration  are being developed constantly \cite{Timofeev2017}. Additional possibilities to such quest are provided by THz detectors based on micro- (MEMS) or nanoelectromechanical (NEMS) systems, especially those employing novel 2D materials \cite{Ferrari2015} such as graphene. Due to the extraordinary mechanical properties, atomically thin NEMS might yield a significant sensitivity  improvement in the operation of mechanical radiation detection devices.

THz detection using graphene has aroused considerable interest \cite{Vicarelli2012,Sensale-Rodriguez2012,Mittendorff2013,Muraviev2013,Cai2014,Wang2014}. Previous work has taken advantage of graphene's linear band structure and the low heat capacity of single-layer graphene.  THz detection has  been done via a plasmonic mechanism \cite{Muraviev2013}, by bolometric detection \cite{Mittendorff2013}, and by noise thermometry \cite{Wang2014}. A recent experiment \cite{Cai2014} with graphene FET with dissimilar contact metals reached noise equivalent power (NEP) around 20 pW/Hz${}^{1/2}$ operating at room temperature. Svintsov et. al. \cite{Svintsov2014} have proposed a scheme with suspended graphene FET, where they take advantage of the plasma resonance that naturally occurs at THz frequencies for short graphene devices. The results so far have remained inferior to the current state-of-the-art bolometers based on superconducting detectors, which reach noise equivalent powers (NEP)  around 10 fW/Hz${}^{1/2}$ in the 0.2 -- 1.0 THz band at $T = 4.2$ K \cite{Penttila2006} and below 1 aW/Hz${}^{1/2}$ at 20 mK \cite{DeVisser2014}. For coherent detectors the relevant figure of merit is the receiver noise temperature $T_n$. For SIS mixer receiver noise temperatures a few times above the quantum limit $hf/2k_B$ (with $h$ and $k_B$ denoting the Planck constant and the Boltzmann constant, respectively) have been reported \cite{Baryshev2015}. Ultra-low-noise. coherent receiver operation of graphene at THz is largely unexplored, albeit graphene based integrated subharmonic mixer circuits have been demonstrated at 200-300 GHz frequencies \cite{Andres2015,Stake2016}.

Here we propose an original scheme of detecting THz radiation using  antenna-coupled mechanical resonators based on atomically thin two-dimensional materials. As the performance of this scheme relies heavily on the properties of the mechanical detector element, we have chosen to employ graphene  in our device. Graphene shows great promise for superior sensitivity owing to its high Young's modulus $ E \sim 1$ TPa and extremely light weight. For optimized radiation detection, the mechanical capacitance has to be matched to a measurement system which is done by employing an electromagnetic cavity (or lumped element circuit). Consequently, our mechanical THz detection setting resembles an optomechanical system, but has a  different type of coupling between electrical signals and the mechanical motion. As in microwave optomechanics, our setup requires superconducting elements to deliver sufficiently large quality factors, which facilitate our detectors  to reach the quantum limit of sensitivity.

\begin{figure}
\includegraphics*[width=0.45\textwidth]{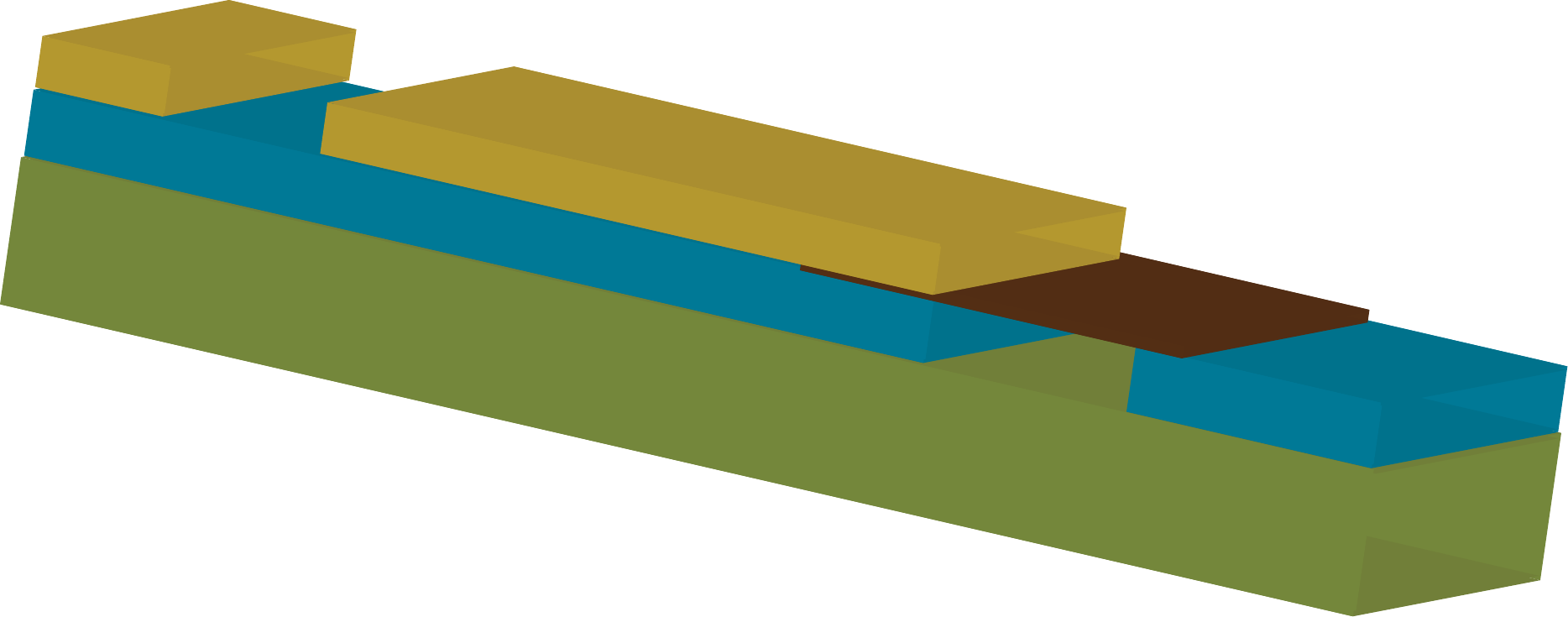}
\includegraphics*[width=0.45\textwidth]{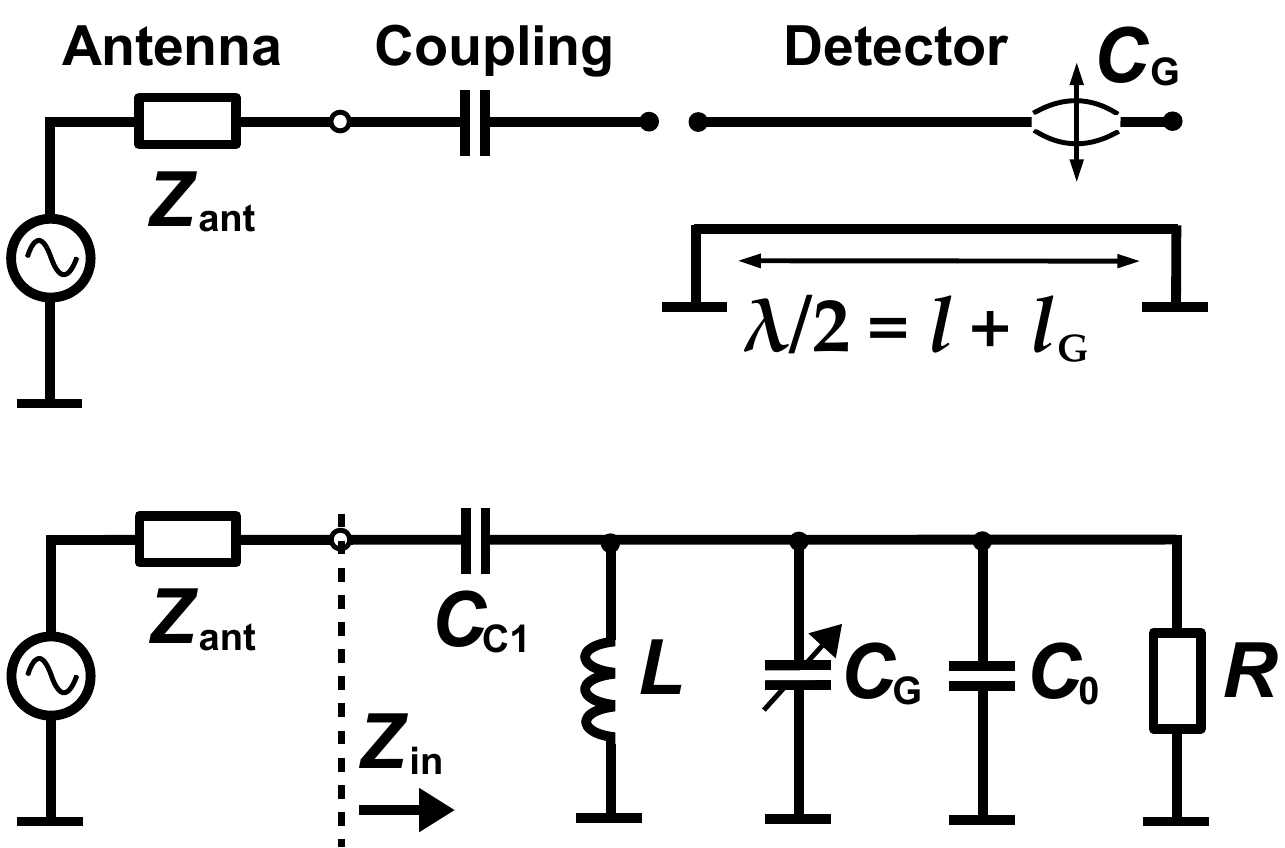}
\caption{ (color on line) a) Schematic view of the analyzed MEMS THz detection system. An atomically thin, suspended membrane, e.g. graphene (brown color), forms a  capacitor located at the end of an electrical cavity resonator line (yellow color), approximately 50 micron long. Antenna is coupled to the MEMS detector via a coupling capacitor formed by the gap in the transmission line. b) Schematic circuit of the radiation detector. The length of the detector unit corresponds to a half-wavelength resonator: $\lambda/2= \ell+\ell_G$ with $\ell_G$ denoting the length of the graphene sensor. c) Equivalent electrical circuit for sensitivity analysis. The equivalent capacitance of the strip line (or coplanar line) electrical resonator is denoted by $C_0$ while the sensor and coupling capacitances are given by $C_G$ and $C_c$, respectively. $R$ denotes electrical losses, mostly caused by the resistance of the two-dimensional material.}
\label{Antenna}
\end{figure}

\section{Power resolution}

We analyze a system where a mechanical resonator is coupled to a THz regime antenna. The signal picked up by the antenna drives the mechanical resonator, and the resulting mechanical vibrations are detected. In our analysis, we don't consider the detection of mechanical motion - we just assume that it can be done down to the oscillation amplitude limit set by  thermal excitation. Demonstration of detection of such thermally driven motion down to 50 mK has been presented, for example, in Ref. \cite{Song2014} in a bilayer graphene NEMS resonator using cavity optomechanics.

Dipole or log spiral antennas can be designed with real impedance of several tens of ohms, the value of which can be matched to a transmission line resonator with a capacitance \textit{C${}_{c}$} by requiring \textit{Z${}_{in}$ = Z${}_{ant}$} (Fig. \ref{Antenna}). In the transmission line case, the input impedance is given by
\begin{equation}
Z_{in,d} =\frac{\pi }{2Q_{e} Z_{0} \left(\omega C_{c} \right)^{2} } ,
\end{equation}
where \textit{Q$_{e}$} and \textit{Z$_{0}$} are the quality factor of the resonator and the characteristic impedance, respectively, and $\omega$ is the angular frequency of the electromagnetic radiation. In Fig. \ref{Antenna}a the geometry is such that the mechanical resonator covers only a fraction of the cavity length at the end of the microwave (THz)  resonator. However, it is also possible in principle that the well-conducting mechanical resonator material forms the electrical cavity by itself. Such a geometry would  be optimal for detection sensitivity (see below).

The antenna can also be matched to a mechanical resonator whose dimensions are smaller than the wavelength, in which case the matching is done by a separate, lumped element LC structure with equivalent performance. In the lumped element case, the matching is done by setting $Z_{in,l} =\frac{1}{Q_{e} \sqrt{L/C} \left(\omega C_{c} \right)^{2} }$ equal to the antenna impedance, \emph{i.e. }$\sqrt{L/C}$ acts as the characteristic impedance $Z_0$ of a corresponding distributed resonator.

The voltage  amplitude of the standing wave induced by a signal can be expressed as
\begin{equation}
V^{2} =Q_{e} \sqrt{\frac{L}{C} } P_{sig},
\label{Eq_signal}
\end{equation}
where the capacitance $C=C_G+C_0$ includes the sensor capacitance $C_G$ and the equivalent capacitance $C_0$ of the strip line (or coplanar line) of the electrical resonator.  The mechanical resonator then feels an eletrostatic force according to $F=\frac{1}{2} \frac{C_GV^{2} }{d}$, where \textit{d} is the vacuum gap between the graphene element and its counterelectrode (ground). Using the two equations above, the force responsivity $\partial$F/$\partial$P$_{sig}$ becomes
\begin{equation}
\frac{\partial F}{\partial P_{sig} } =\frac{Q_{e} \sqrt{LC} }{2d} \sqrt{\frac{C_G}{C_0+C_G}}.
\end{equation}
 In the ideal case, when electrical cavity is formed by the mechanical resonator, the capacitive shunting factor $\sqrt{\frac{C_G}{C_0+C_G}}$ reducing the sensitivity becomes equal to one, while for the setting in Fig. \ref{Antenna}a it equals to $\sim 0.3$ ($C_G \simeq 0.1 C_0$).

 The power resolution of the detector is ultimately limited by  mechanical thermal noise. The mechanical noise can be expressed using the
 fluctuation-dissipation theorem as $S_{F}^{1/2} =\sqrt{\frac{2}{\pi} k_B T \gamma_m}$ where  \textit{T } is the temperature of the detector element (phonon temperature), and $\gamma_m$ denotes the damping rate of the mechanical resonator; at low temperatures electronic temperature may deviate substantially from phonon temperature due to weak electron-phonon coupling \cite{Laitinen2014}. By replacing  the decay rate $\gamma_m/2\pi= \frac{\omega _{m} m}{Q_{m}}$, the force fluctuations can be written as
 $S_{F} ^{1/2} =\sqrt{\frac{4k_B T \omega _{m} m}{Q_{m}}}$ , where $\omega_{m}$ is the mechanical angular frequency, \textit{m} denotes the resonator mass \footnote{We neglect here the mechanical mode shape which could lower the effective mass down to $0.4 \times m$.}, and \textit{Q${}_{m}$} is the mechanical quality factor. Noise equivalent power is then given by $NEP=\frac{S_{F} ^{1/2} }{\partial F/\partial P_{sig} }$, which leads to
\begin{equation}
NEP=\frac{2d\omega }{Q_{e} } \sqrt{\frac{4k_{B} T\omega_{m} m}{Q_{m} } } \frac{C_0+C_G}{C_G}.
\label{NEP}
\end{equation}
The above equation  works generally for any antenna matched MEMS system, regardless of the design details. Thus, it can be used to estimate performance of the system, if the characteristics of the electrical and mechanical resonators are known. The only requirement is that the resonator motion can be measured at the thermal motion level. Although the detector can work at room temperature, lowering the temperature directly improves sensitivity by reducing thermal noise as shown in Eq. \eqref{NEP}, and typically the $Q_m$ values of mechanical graphene resonators are significantly larger at cryogenic temperatures.

A substantial amount of experiments have been performed on graphene mechanical resonators \cite{Bunch2007,Chen2009,Song2012,Song2014,BachtoldNL2014,Singh2014} and hence we can reliably estimate the expected performance. For a realistic case study, we adopt the experimental parameters for a bilayer graphene resonator from Ref. \onlinecite{Song2014}, i.e. $\omega $${}_{m}$/2$\pi $ = 24 MHz, vacuum gap $d$ = 70 nm, and \textit{Q${}_{m}$} = 15~000, and $m$ = 10$^{-17}$ kg. Operated at $T=$ 25 mK, the NEP goes down to $\sim$ 1.3 fW/Hz${}^{1/2}$, provided that the matching circuit can reach $Q_{e} \frac{C_G} {C_0+C_G}= 100$ at 500 GHz operating frequency. If the temperature is increased to 3 K, which can be reached with a pulsed cryocooler, the NEP increases to $\sim$ 14 fW/Hz${}^{1/2}$. Here we assume that the practical $Q_m$ for a graphene mechanical resonance at 3 K  is not altered significantly from 15~000. The bandwidth of the device is only around $\Delta $$\nu $ = $1/Q_e$ i.e. a few gigahertz, which should be contrasted to the broadband detection $\Delta $$\nu $ = 500 GHz offered by bolometric detectors. Hence, no improvement compared with present techniques is achieved by our mechanical detection scheme. This result is quite expected as the high impedance of the small sensor element allows only a narrow-band detection to be utilized.

%

\section{Coherent detection}

The device analyzed above can also be used as a mixer, in which case its operation can be brought down to the ultimate sensitivity limit governed by the Heisenberg uncertainty principle.
For a couple of decades, mixers based on superconductor-insulator-superconductor (SIS) junctions or superconducting hot electron bolometers (SHEB) have formed the main state-of-the-art at frequencies up to few THz. Reliable quantum-limited mixers can be achieved at frequencies below 680 GHz \cite{Tucker1985}, a limit which is set by the superconducting gap of Nb. At frequencies $f > 4\Delta/h$, where $2\Delta$ is the Cooper pair breaking energy, significant losses are introduced in the superconductor. While this does not prohibit heterodyne mixing at larger frequencies, the noise temperature is then typically limited to few times the quantum limit $T_q$ \cite{Bin1996, Jackson2001, Jackson2005, Westig2013}.


A local oscillator (LO) signal \textit{V}${}_{LO}$ applied over the MEMS structure, is summed with the measured signal \textit{V}, so that the RMS force acting on the resonator at frequency $\omega -\omega_{LO}$ is given by
\begin{equation}
F\left(\omega -\omega _{LO} \right)=\frac{C_G}{\sqrt{2} d} VV_{LO} .
\end{equation}
Voltage noise can be written as
\begin{equation}
S_{V} ^{1/2} =\frac{1}{\partial F/\partial V} S_{F} ^{1/2} =\frac{\sqrt{2} d}{C_G V_{LO} } \sqrt{\frac{4k_{b} T\omega _{m} m}{Q_{m} } } \left(\frac{C_0+C_G}{C_G} \right).
\end{equation}
By using Eq. \eqref{Eq_signal}, the noise power can be referred to the signal power by $S_{P}= S_{V} / \left[Q_{e}(L/(C_0+C_G)^{1/2} \right]$. By defining the noise temperature as $T_{n} = S_{P}/k_{b}$,
\begin{equation}
T_{n} =\frac{2d^{2} \omega }{k_{b} CV_{LO} ^{2} Q_{e} } \left(\frac{4k_{b} T\omega _{m} m}{Q_{m} } \right).
\end{equation}
We further replace $V_{LO}$ with $P_{LO}$  in Eq. \eqref{Eq_signal}, and rewrite
\begin{equation}\label{Tn}
T_{n} =\frac{2d^{2} \omega ^{2} }{P_{LO} Q_{e} ^{2} } \left(\frac{T\omega _{m} m}{Q_{m} } \right) \left(\frac{C_0+C_G}{C_G} \right)^2.
\end{equation}
We emphasize that the quoted noise temperature  contains only the noise fluctuations stemming from the mechanical dissipation. There is another noise contribution due to electrical fluctuations which are limited to zero point fluctuations in the signal band since $\hbar\omega \gg k_B T$. The noise power per unit band for quantum fluctuations in a cavity mode at frequency $\omega/2\pi$ corresponds to an energy of half of a photon: $\frac{1}{2}\hbar\omega$. Under impedance matching conditions for signal frequency, this corresponds to the single-side-band noise temperature of $T_n = \hbar\omega/\left(2k_B\right)$, i.e. the standard quantum limit. In practice, however, $T_n = \hbar\omega/\left(k_B\right)$ since separation of the image frequency is problematic in the THz regime, which leads to additional quantum noise of $\frac{1}{2}\hbar\omega$.

In order to reach the quantum limit in detection sensitivity, the contribution of the mechanical fluctuations needs to be brought below that of the quantum noise. Eq. \eqref{Tn} suggests that this is accomplished by increasing $P_{LO}$ sufficiently. A fundamental limitation for $P_{LO}$ arises from the linearity requirement for the mechanical resonator motion. Furthermore, practical limitations include that the LO power dissipated in a dilution refrigerator at  20 mK has to be limited to about 20 $\mu$W. Using Eq. \eqref{Tn}, and the experimental graphene resonator and matching circuit parameters quoted in Section II, we find that the LO power of about 100 nW, well in line with typical cryostat operation, is sufficient for reaching the quantum limit of $T_n\approx$ 12 K at $\omega/2\pi$ = 500 GHz. We have also checked that the excitation at $ f_m$ via mixing (cf. drive e.g. in Ref. \onlinecite{Song2012}) and at $f_{LO}$ ($\gg f_m$) remain well in the linear regime of graphene mechanical motion. It is worth to note that, according to Eq. \eqref{Tn}, the noise temperature scales down with the resonator mass provided that other parameters can be kept unchanged. Hence, narrow ribbons are expected to yield the optimum, but the optimal width will depend strongly on other constraints like the capacitive participation ratio $\frac{C_G}{C_0+C_G} $.

There are two critical issues concerning practical applications: 1) Dissipation in electrical cavity at THz frequency, i.e. the value of $Q_e$, and 2) Participation ratio of the graphene sensor capacitance.
High $Q_e$-values have been reported to whispering gallery modes at THz frequencies \cite{Campa2015}. If large-gap superconducting materials can be employed for on lithographic chip circuits, there are no principal obstacles in obtaining $Q_e \sim 10^4$ up to 1 THz or 2.5 THz using NbTiN or MgB$_2$, respectively \cite{Brandstetter2012}. Such a $Q_e$-factor combined with a graphene participation ratio a few per cent would bring the thermal noise contribution well below the quantum noise. The performance would improve even further if $R_{\square} < 10$ $\Omega$ can be employed as targeted for graphene touch screen displays  \cite{Wassei2010}.

Our analysis has not discussed practical problems in the detection of the vibration of the graphene membrane. As the participation ratio of the graphene capacitance becomes larger, heating by Joule dissipation becomes stronger. However, the electron-phonon coupling at mK temperatures is weak and the environmental temperature of the first fundamental mode grows up moderately with heating of the electrons, provided the Kapitza resistance between graphene and its support structure is optimized \cite{Song2014}. Hence, operation at the necessary high drive powers is  feasible in our detector configuration without losing sensitivity to imposed radiation.

\section*{Acknowledgements}
This work was supported by the Academy of Finland (grant no. 286098 and 284594, LTQ CoE), and FP7 FET OPEN project iQUOEMS. This research made use of the OtaNano Low Temperature Laboratory infrastructure of Aalto University, that is part of the European Microkelvin Platform.

\end{document}